\documentclass[prd,amsmath,floats,amssymb, floatfix,superscriptaddress,nofootinbib,onecolumn]{revtex4} 
\usepackage{multirow}

\usepackage[plainpages=false, colorlinks=true, anchorcolor=blue, linkcolor=blue, citecolor=blue, bookmarks=false]{hyperref}
\usepackage[T1]{fontenc}
\usepackage{graphicx}
\usepackage{subcaption}
\usepackage{dcolumn}
\usepackage{bm}
\usepackage{adjustbox}
\usepackage{booktabs} 
\newcommand{\rthis}[1]{\textcolor{black}{#1}}
\usepackage{float}
\usepackage[paperwidth=210mm,paperheight=297mm,centering,hmargin=2cm,vmargin=2.5cm]{geometry}

\begin{document}
\include{notations}
\preprint{APS/123-QED}

\title{Search for MeV Gamma-ray emission from TeV bright red dwarfs with COMPTEL}

\author{Niharika Shrivastava}
 \altaffiliation{Email:niharikas21@iiserb.ac.in}
 \affiliation{Department of Physics, Indian Institute Of Science Education and Research, Bhopal, Madhya Pradesh, 462066, India}

 \author{Siddhant Manna}
 \altaffiliation{Email:ph22resch11006@iith.ac.in}
 
\author{Shantanu Desai}
 \altaffiliation{Email:shntn05@gmail.com}
\affiliation{Department of Physics, IIT Hyderabad, Kandi, Telangana 502284,  India}

\begin{abstract}
The SHALON atmospheric Cherenkov telescope  has detected very high energy gamma-ray emission at TeV energies from eight red dwarfs, namely, 
V388 Cas, V547 Cas, V780 Tau, V962 Tau, V1589 Cyg, GJ 1078, GJ 3684 and GL 851.1. Consequently, these red dwarfs   have been suggested as sources of ultra-high energy cosmic rays. 
In this work, we search for soft gamma-ray emission from  these  TeV bright red dwarfs between 0.75-30 MeV using archival data from the COMPTEL gamma-ray imaging telescope, as a follow-up to a similar search for GeV gamma-ray emission using the Fermi-LAT telescope.  Although,  prima-facie, we detect  non-zero  photon flux from three red dwarfs with high significance, these signals  can attributed to contamination from  nearby sources such as Crab and Cygnus, which are within the angular resolution of  COMPTEL, and have been previously detected as very bright point sources at MeV energies. Therefore, we could not detect any statistically significant signal ($>3\sigma$) from any of these eight  red dwarfs from 0.75-30 MeV.  We then report the 95\% confidence level upper limits on the differential  photon flux (at 30 MeV),  integral photon  flux  and integral energy flux for all of the eight  red dwarfs. The integral energy flux limits range between $10^{-11}-10^{-10} \rm{ergs/cm^2/s}$.
\end{abstract}

\keywords{}

\maketitle
\section{\label{sec:level1}Introduction\protect}
Red dwarfs are one of the smallest stars on the main sequence with masses between 0.075-0.5 $M_{\odot}$ and surface temperatures between 2,500-5,000 K~\cite{Edgeworth}. They are sometimes synonymous with stellar M-dwarfs. These stars  frequently emit flares characterized by a power law with index between 1.7 and 2.4~\cite{Aschwanden}, where  the energy of each flare is  between $10^{32}$ and $10^{35}$ ergs~\cite{Yang17}. They account for more than 70\% of the galactic stars. 

Energetic bursts from red dwarfs have been detected throughout the electromagnetic spectrum.
The SWIFT-BAT telescope detected a  hard X-ray  outburst  between 15-50 keV from the flaring dwarf DG CVn, with a luminosity of $1.9 \times 10^{32}$ ergs/sec along with associated optical emission~\cite{Drake}. A coincident radio flare with flux density greater than 100 mJy  was also detected with the AMI-LA radio telescope at 15 GHz~\cite{Fender}. However, no associated emission in gamma-rays between 0.1-100 GeV was seen by  the Fermi-LAT detector during this flare~\cite{Loh}. Nevertheless, a gamma-ray pulse from another red dwarf, namely  TVLM 513-46546 was detected by Fermi-LAT with a power law index of $2.59 \pm 0.22$ and having   the same periodicity as that seen at optical wavelengths~\cite{Song}. Most recently, very high energy gamma-ray emission between 800 GeV to 20 TeV  has been detected from eight red dwarfs by the SHALON atmospheric Cherenkov telescope~\cite{Shalon}. Some of these red dwarfs have also been detected in X-rays by the ROSAT satellite~\cite{Gershberg}.
For all these reasons, red dwarfs have also been proposed as sources of ultra high energy cosmic rays~\cite{rdcosmicrays}. Motivated by all these considerations, a search for GeV gamma-rays from these eight red dwarfs was done using 13.6 years of  Fermi-LAT in the energy range from 0.2-500 GeV-~\cite{Huang24}. Although GeV gamma-ray emission was seen in the vicinity of two red dwarfs, this signal was excluded as emanating from these red dwarfs based on energetic arguments.
Consequently, no significant emission associated with any of the eight  red dwarfs could be found~\cite{Huang24}.

In this work, we look for gamma-ray emission at MeV energies using legacy data from the COMPTEL gamma-ray telescope on the COMPTON gamma-ray observatory, using the same methodology as our previous work, where we looked for MeV emission from galaxy clusters~\cite{Manna} (M24 hereafter). This manuscript is structured as follows. We discuss the data analysis in Sect.~\ref{sec:level2} and results in Sect.~\ref{sec:results}. Theoretical implications of our results can be found in Sect.~\ref{sec:implications}. We conclude in Sect.~\ref{sec:conclusions}.

\section{Data Analysis}
\label{sec:level2}
The COMPTEL telescope was one of the three instruments onboard the Compton Gamma Ray Observatory (CGRO), \rthis{which}  launched in  April 1991 which took data until June 2000. This telescope had a field of view of about one steradian and  was sensitive to gamma-rays between 0.75- 30 MeV. Its   energy and angular resolution ranged between 5-8\% and $(1.7-4.4)^{\circ}$, respectively depending on the photon energy~\cite{Comptel93}. COMPTEL did about 340 distinct pointings during its nine years of data taking,  where each pointing had a field of view radius of around $30^{\circ}$~\cite{Strong}, where ``field of view radius'' refers to the size of COMPTEL's field of view for a given pointing. More details on the observing specifications and performance of COMPTEL can be found in ~\cite{Comptel93}. COMPTEL has detected a large number of astrophysical sources, such as   pulsars, AGNs, X-ray binaries, gamma-ray bursts, solar flares,  supernova remnants, extragalactic diffuse gamma-ray background, etc~\cite{Comptel93,Comptel00}. No other telescope has imaged the universe in the aforementioned energy range after the decommissioning of CGRO. We also note that results from the full survey mission have not yet been released, and there could be other undiscovered sources in the dataset~\cite{Strong}.
Most recently, new software based on the GammaLib and ctools libraries~\cite{ctools} has been developed facilitating seamless  analysis of the COMPTEL data~\cite{Knod}.  Therefore, this legacy data provides a unique opportunity to probe the uncharted territory of the universe  in soft gamma-rays at  MeV energies.

We mine this data to look for MeV emission from eight red dwarfs, which have been detected by SHALON at TeV energies. We basically follow the same analysis methodology as M24. We carried out  a systematic search within a $30{^\circ}$ radius surrounding each target red dwarf utilizing the revamped COMPTEL analysis framework implemented in the ctools software package~\cite{Knod}. During the nine years of satellite operations, the CGRO mission was divided into nine phases and each of these phases had multiple viewing periods, with  each having a single pointing direction. These so-called viewing periods (vp) were typically of 14 days duration. We queried all the COMPTEL  pointings encompassing all the nine phases of viewing for all the red dwarfs.  The number of viewing periods for each red dwarf are listed in  Table~\ref{tab:Table_TS}. There was one particular viewing period, vp.0426 (phase 04), pointing to the Galactic anticenter, which was unreadable by the software due to corrupt event processing file and is thereby excluded from our analysis. It was within a $30^\circ$ radius of three of our red dwarfs, namely V780 Tau, V962 Tau and GJ 1078.

Before commencing the COMPTEL data analysis, we first generated a database from the data available in the HEASARC archive using the \texttt{comgendb} tool. Subsequently, we employed the \texttt{comobsselect} tool to extract the relevant viewing periods for each red dwarf from the COMPTEL database hosted by HEASARC. For a targeted analysis of each energy bin, the \texttt{comobsbin} tool was used to bin the observations into 16 logarithmically spaced energy bins from 0.75 to 30 MeV. \rthis{The binning was chosen based on} COMPTEL's energy resolution.

COMPTEL data is analyzed in a 3-dimensional data space, \rthis{hereafter referred to as the
Compton data space,  consisting of the} Compton scatter angle and the Compton scatter direction; see ~\cite{Knod} for more details.
Given the multiple viewing periods, we utilized the \texttt{comobsadd} tool with 80 bins each with bin size of 1${^\circ}$ in both the  Compton scattering directions, while the Compton scattering angle had 25 bins each with a bin size of 2${^\circ}$. We chose the default values as recommended with the software. This tool concatenates individual viewing periods into a  single  bin in the Compton data space for each energy band, significantly enhancing the speed of the analysis compared to conducting a joint analysis of each viewing period individually. The cumulative effective exposure from all the viewing periods for all the red dwarfs is presented in Table~\ref{tab:Table_TS}.

For maximum likelihood analysis on the data, we generated the model definition file using the \texttt{comobsmodel} tool. This file facilitated fitting the background model to the data and enabled us to analyze our red dwarfs using different templates for point and diffuse sources. Details about the background model validation can be found in ~\cite{Knod}.
We employed point source, radial disk (radius = $0.2{^\circ}$), and radial Gaussian templates ($\sigma$ = $0.2{^\circ}$) as outlined in ~\cite{Wood2017}. For all practical purposes, we would expect the signal to be a point source. Nevertheless, for completeness we also analyzed using the other two extended templates.

The spectral analysis was conducted using the \texttt{csspec} tool, akin to the methodology used in M24. 
This tool computes the source spectrum by fitting a model within a given set of spectral bins. It also calculates an upper flux limit, which is particularly useful when the source is not significantly detected within a spectral bin. For this analysis, we applied the \texttt{BINS} method, where the spectral model is replaced by a  function that fits  the data separately in each bin. More details  about the \texttt{BINS} method can be found in the ctools documentation~\cite{ctools}\footnote{\url{http://cta.irap.omp.eu/ctools/users/reference_manual/csspec.html}}.

To evaluate the significance we calculated the following test statistic~\cite{Mattox1996}:
\begin{equation}
\rm{TS} = 2 \ln L(M_s + M_b) - 2 \ln L(M_b),
\label{eq:TS}
\end{equation}
where $\ln L(M_s + M_b)$  represents   the likelihood when both the source model ($M_s$) and the background model ($M_b$) are  fitted to the data. According to Wilks' theorem, TS follows a $\chi^{2}$ distribution with $n$ degrees of freedom, where $n$ is equal to  the number of free parameters for the source model~\cite{Wilks}. This same statistics is also used in analysis of data from Fermi-LAT and IceCube~\cite{MannaFermi,Pasumarti,Pasumarti2}. The detection significance or $Z$-score is given by the square root of TS. Therefore, for a 3$\sigma$ (5$\sigma$) detection, TS has to be greater than 9 (25). Smaller values  of TS implies that the data is consistent with background.  In case of null detection,  we presented the upper limits for the differential photon flux  using a reference energy of 30 MeV, similar to M24.

\section{Results}
\label{sec:results}
Our results for the  TS values of all the eight red dwarfs can be found in Table~\ref{tab:Table_TS}. 
The reported TS values are the mean values of the TS across all the 16 energy bins. The TS for each energy bin was calculated using the \texttt{csspec} tool which we used to make the Spectral Energy Distribution (SED) plots.

\begin{figure}[htbp]
    \centering
    \begin{subfigure}{0.45\textwidth}
        \centering
        \includegraphics[width=\textwidth]{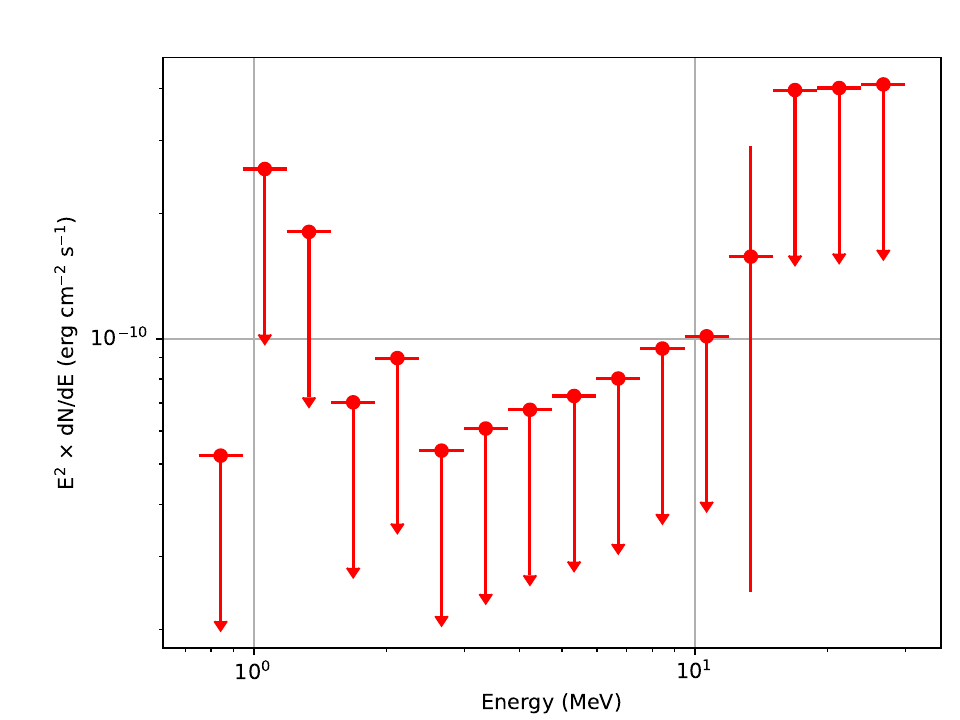}
        \caption{V388 Cas}
        \label{fig:RD1}
    \end{subfigure}\hfill
    \begin{subfigure}{0.45\textwidth}
        \centering
        \includegraphics[width=\textwidth]{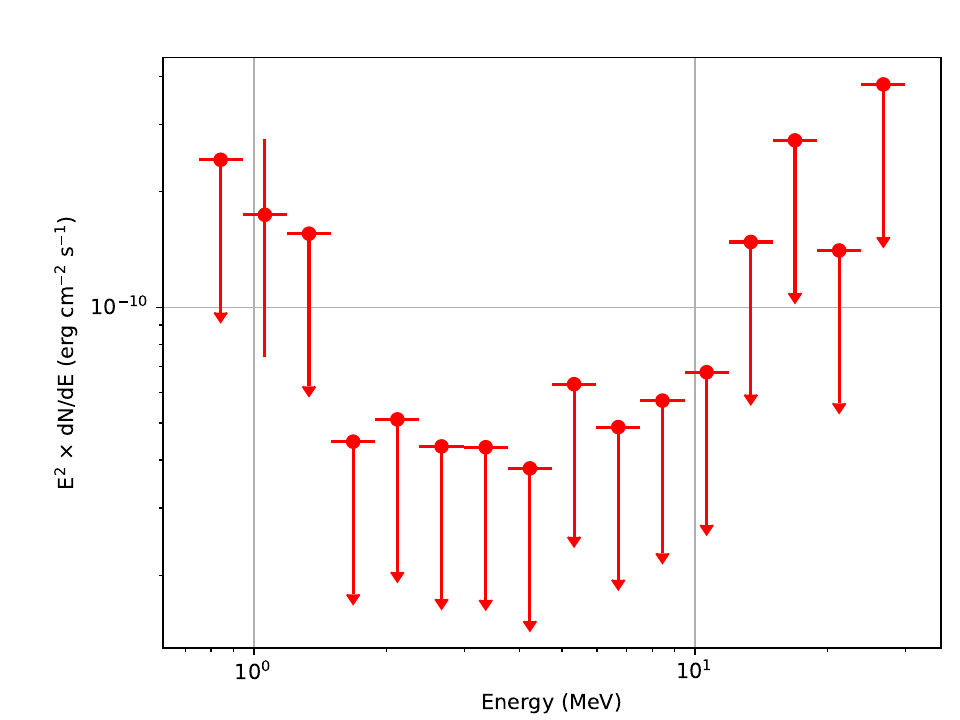}
        \caption{V780 Cas}
        \label{fig:RD2}
    \end{subfigure}
    
    \vspace{1em}
    
    \begin{subfigure}{0.45\textwidth}
        \centering
        \includegraphics[width=\textwidth]{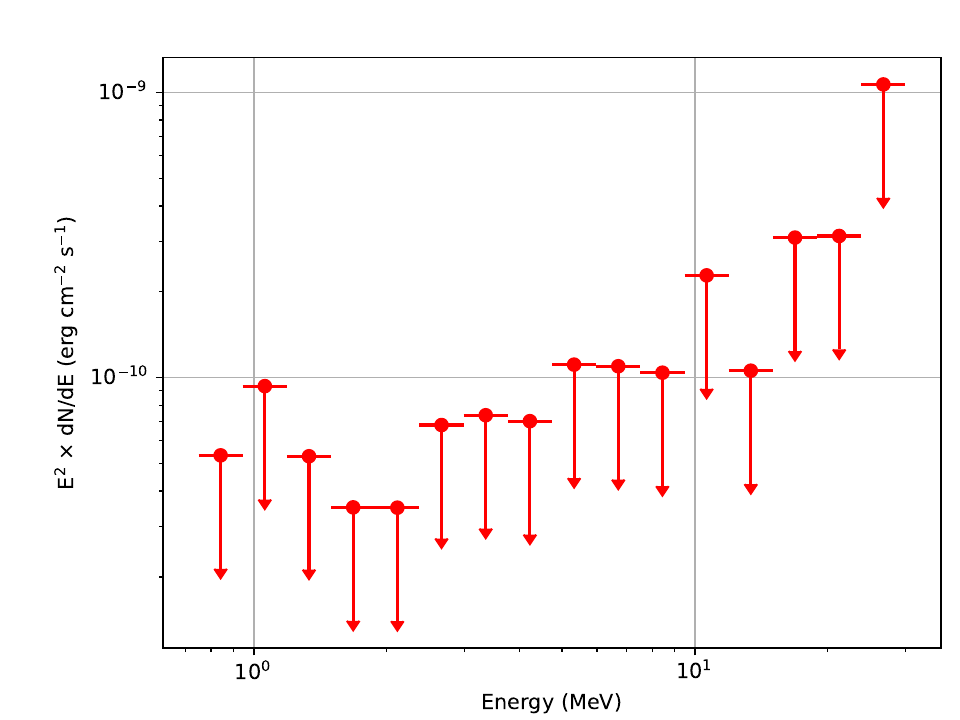}
        \caption{ GJ 3684}
        \label{fig:RD7}
    \end{subfigure}\hfill
    \begin{subfigure}{0.45\textwidth}
        \centering
        \includegraphics[width=\textwidth]{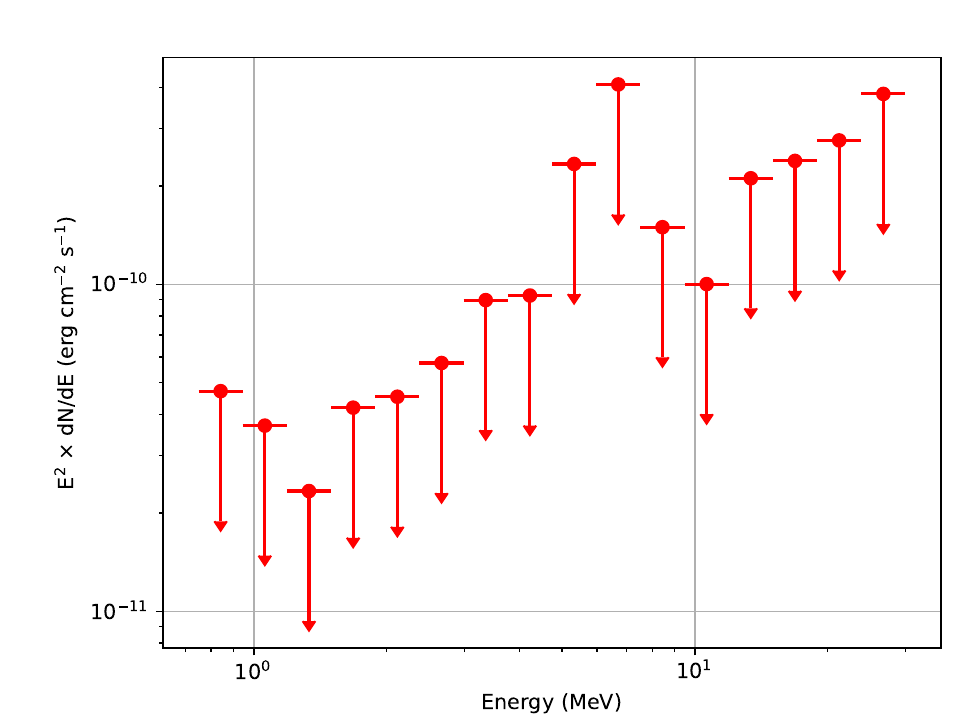}
        \caption{ GL 851.1}
        \label{fig:RD8}
    \end{subfigure}
    
    \caption{SED plot using single point source analysis of red dwarfs. The downward pointing arrows refer to 95\% c.l. upper limits.}
    \label{firstfig}
\end{figure}

\begin{figure}[ht!]
    \centering
    \begin{subfigure}[b]{0.45\textwidth}
        \centering
        \includegraphics[width=\textwidth]{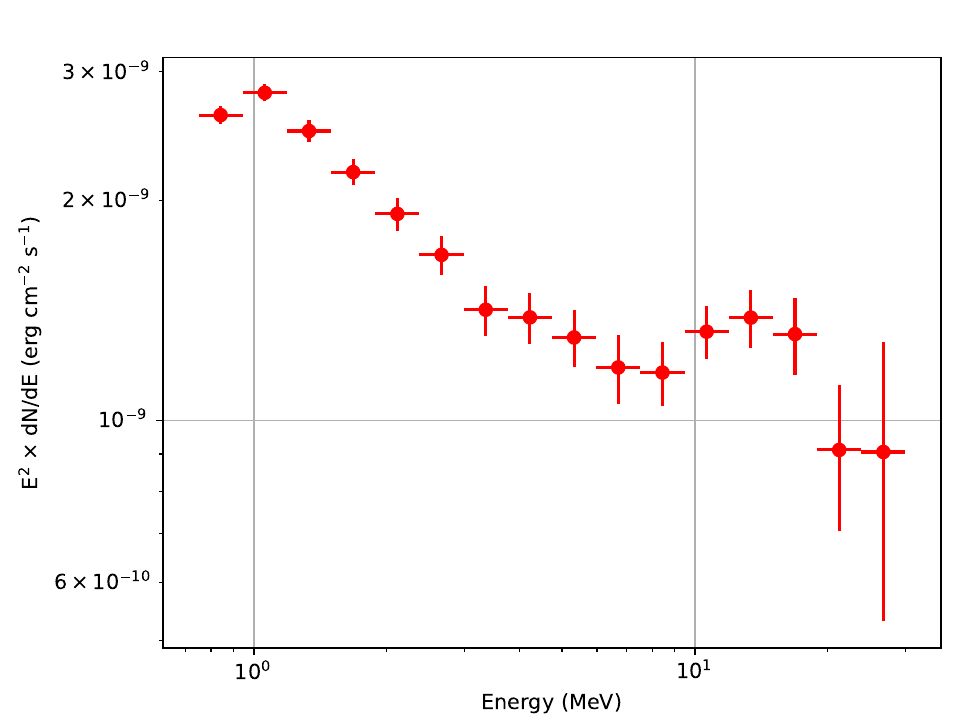}
        \caption{SED using a single point source template }
        \label{fig:sub5}
    \end{subfigure}
    \hfill
    \begin{subfigure}[b]{0.45\textwidth}
        \centering
        \includegraphics[width=\textwidth]{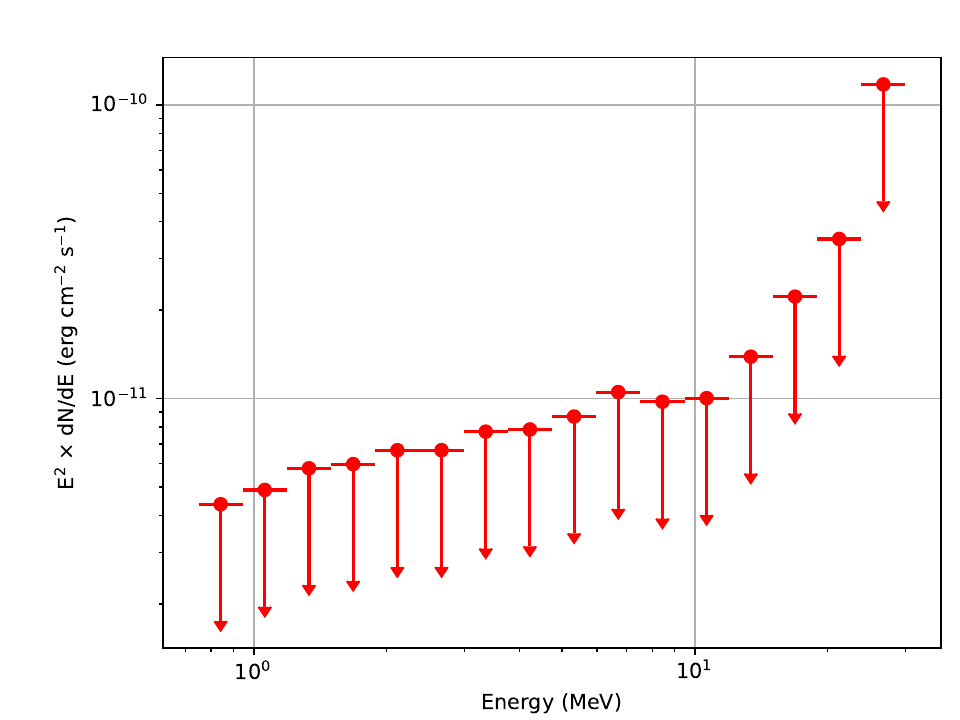}
        \caption{SED using two point source templates   including  the Crab pulsar.}
        \label{fig:sub6}
    \end{subfigure}
    \caption{One point and two point source  analyses plots for V780 Tau. The downward pointing arrows refer to 95\% c.l. upper limits.}
    \label{fig:RD3}
\end{figure}

\begin{figure}[ht!]
    \centering
    \begin{subfigure}[b]{0.45\textwidth}
        \centering
        \includegraphics[width=\textwidth]{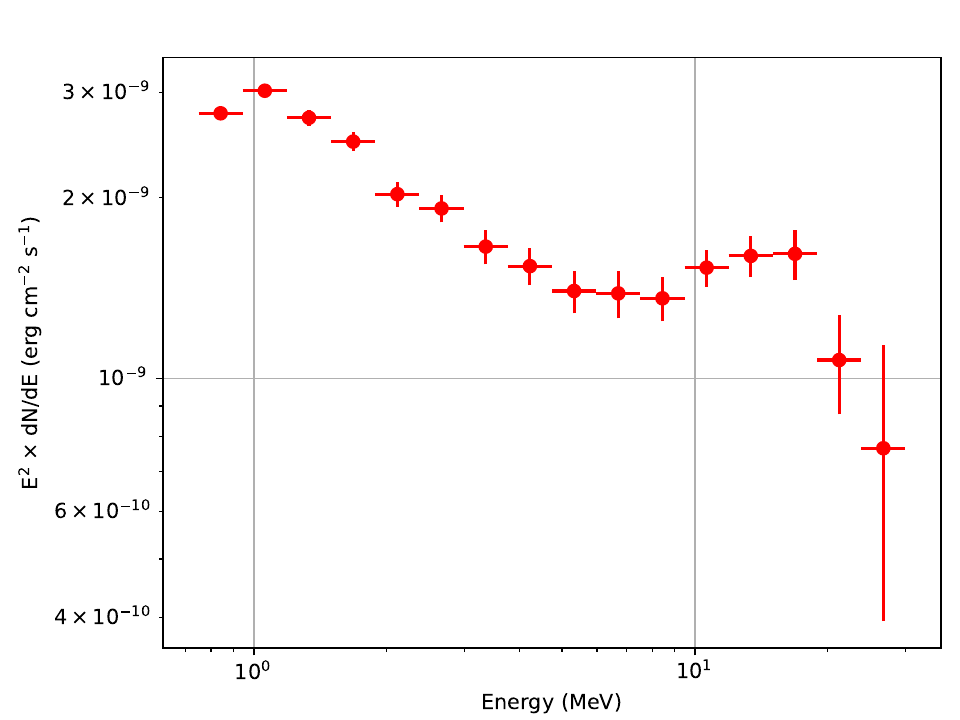}
        \caption{SED using using a single point source template. }
        \label{fig:sub7}
    \end{subfigure}
    \hfill
    \begin{subfigure}[b]{0.45\textwidth}
        \centering
        \includegraphics[width=\textwidth]{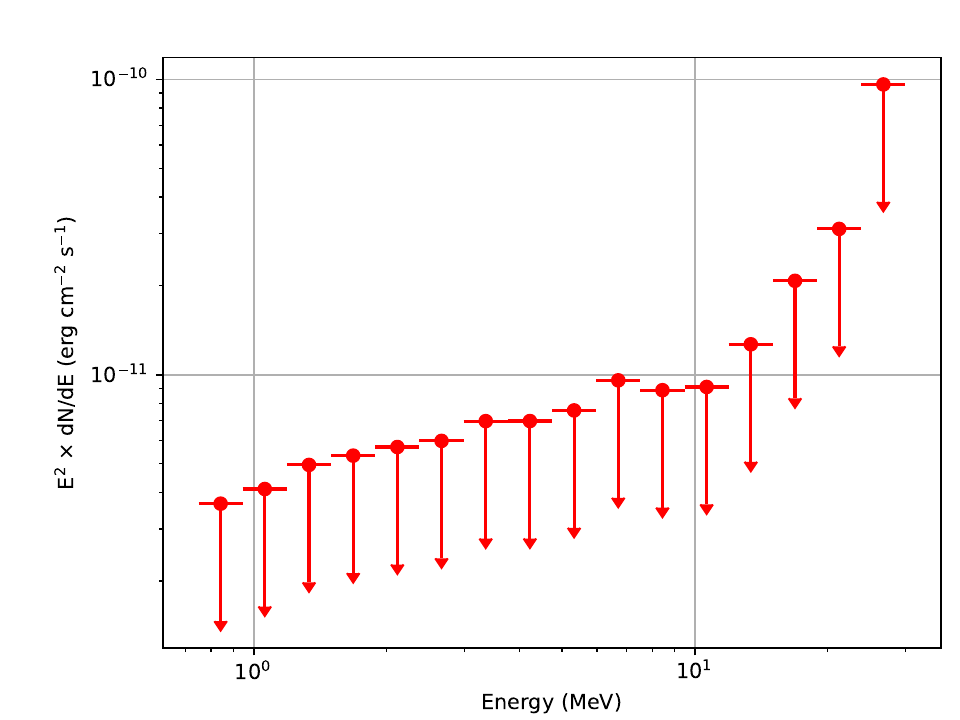}
        \caption{SED using two point source templates   including  the Crab pulsar.}
        \label{fig:sub8}
    \end{subfigure}
    \caption{One point and two point source analyses plots for V962 Tau. The downward pointing arrows refer to 95\% c.l. upper limits.}
    \label{fig:RD4}
\end{figure}

\begin{figure}[ht!]
    \centering
    \begin{subfigure}[b]{0.45\textwidth}
        \centering
        \includegraphics[width=\textwidth]{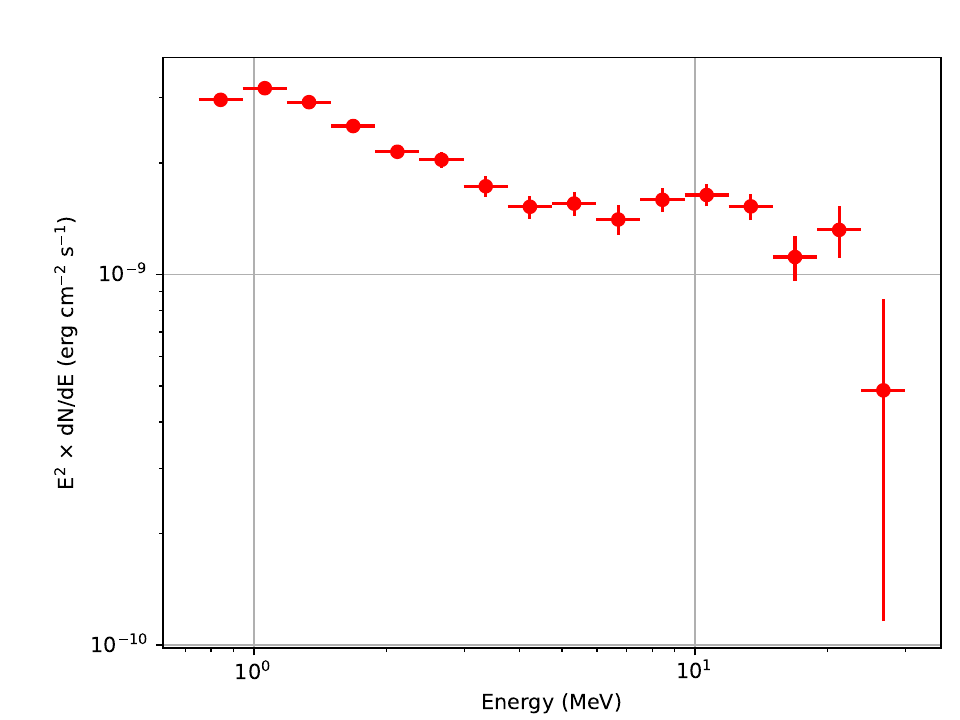}
        \caption{SED using a single point source template.}
        \label{fig:sub11}
    \end{subfigure}
    \hfill
    \begin{subfigure}[b]{0.45\textwidth}
        \centering
        \includegraphics[width=\textwidth]{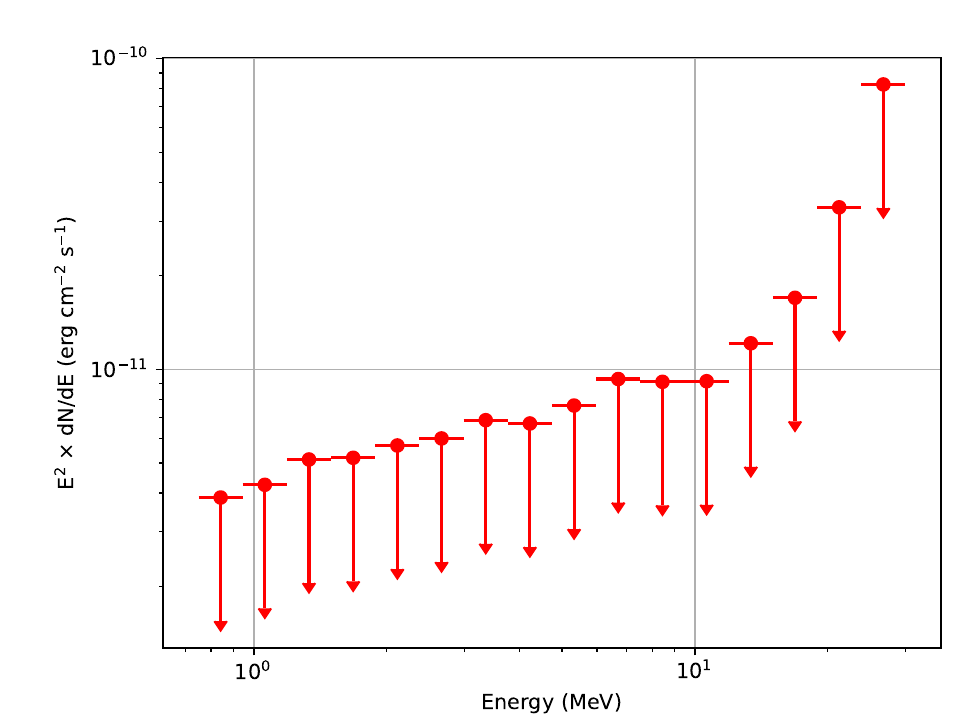}
        \caption{SED using two point source templates   including  the Crab pulsar.}
        \label{fig:sub12}
    \end{subfigure}
    \caption{One-point and two-point  source analyses plots for GJ~1078. The downward pointing arrows refer to 95\% c.l. upper limits.}
    \label{fig:RD6}
\end{figure}

\begin{figure}[ht!]
    \centering
    \begin{subfigure}[b]{0.45\textwidth}
        \centering
        \includegraphics[width=\textwidth]{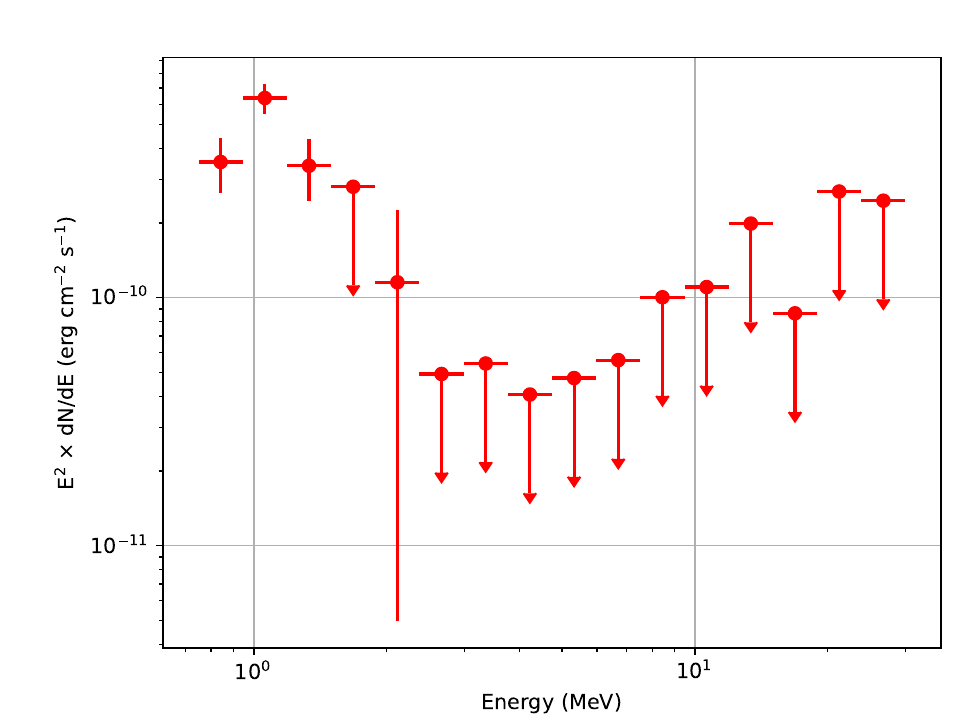}
        \caption{SED using a single point source  template.}
        \label{fig:sub9}
    \end{subfigure}
    \hfill
    \begin{subfigure}[b]{0.45\textwidth}
        \centering
        \includegraphics[width=\textwidth]{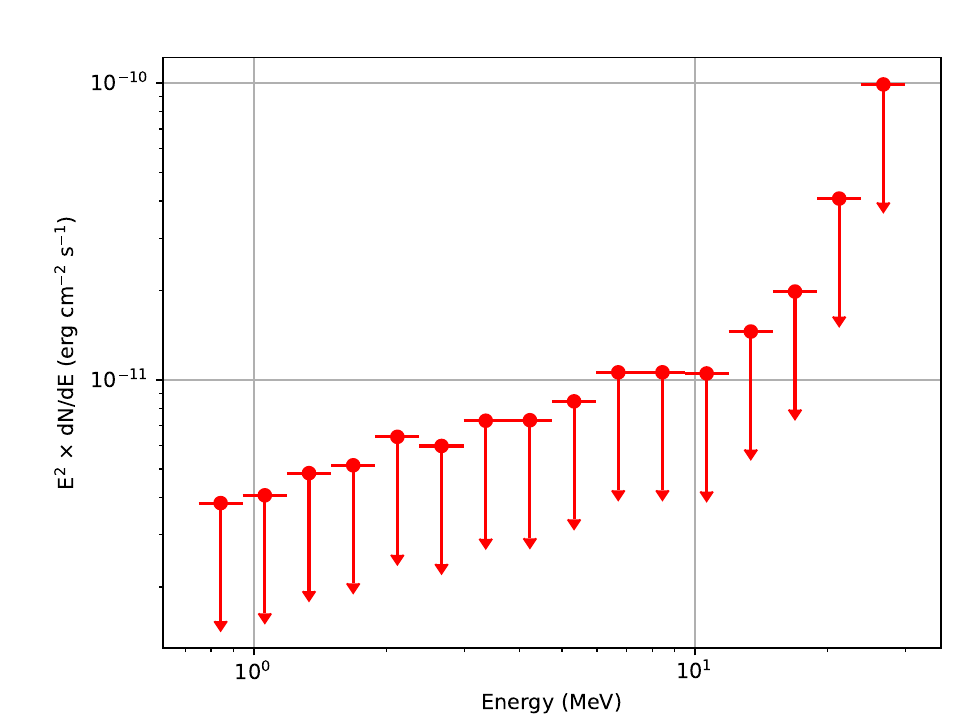}
        \caption{SED using point source templates   including Cygnus.}
        \label{fig:sub10}
    \end{subfigure}
    \caption{One-point and two-point analysis plots for V1589 Cyg}
    \label{fig:RD5}
\end{figure}

The SED plots for all the red dwarfs are shown in Figs.~\ref{firstfig}--\ref{fig:RD5}.  For V388 Cas  we find one bin with non-zero flux at energies of 13.39 MeV (cf. Fig~\ref{fig:RD1})  with significance of 0.59$\sigma$  with a TS value of 1.4, using a point source template. Similarly, for V547 Cas, we find one bin with non-zero flux at energies of 1.06 MeV (cf. Fig~\ref{fig:RD2})  with significance of  1.74$\sigma$  with a TS value of 3.1 using the point source template. Since the significance is less than $3\sigma$ and detected in only energy bin, we consider this a statistical fluctuation and not a real signal. For GJ~3684 and GL~851.1, we report null detections at all energies.

For the remaining two other red dwarfs namely,  V780 Tau and V962 Tau, we detect non-zero flux at the lowest energies from 0.75- 1 MeV with very high significance,  and upper limits beyond that with all the three search templates.
Similarly, GJ 1078 shows a statistically significant flux at all energies. However, these sources are within $(3-4)^{\circ}$ of  the Crab pulsar (cf. Table~\ref{tab:Table_TS}). The angular resolution of COMPTEL gets degraded at low energies making it susceptible to contamination from nearby sources within $4^{\circ}$ at very low energies (J\"urgen Kn\"odlseder, private communication). Among these sources, GJ~1078 is within $0.64^{\circ}$ of the Crab pulsar, which is less than the COMPTEL PSF at all energies~\cite{Comptel93}.
The Crab is the brightest source detected  in MeV gamma-rays, whose flux is  $(16.6 \pm 3.0) \times 10^{-5} \text{ photons cm}^{-2} \text{ s}^{-1}$ between 0.75-1 MeV,  $(3.86 \pm 2.12) \times 10^{-5} \text{ photons cm}^{-2} \text{ s}^{-1}$ between 1-10 MeV, and  $(15.5 \pm 0.5) \times 10^{-5} \text{ photons cm}^{-2} \text{ s}^{-1}$ between 10-30 MeV~\cite{Comptel00}. Therefore, in order to circumvent this issue, for the three aforementioned red dwarfs, we used a template consisting of two sources: one for the Crab pulsar and the other for the red dwarf. Using this two-point source template, we do not detect any signal   at  these energies, and the SED plots  display  only  upper limits at all energies. The SED plots with both the single and two-point source templates can be found in Fig.~\ref{fig:RD3}, Fig.~\ref{fig:RD4}, and Fig.~\ref{fig:RD6}. We find similar results for these red dwarfs  when using a single radial disk or radial Gaussian template and then doing a analysis with two radial disk or radial Gaussian templates.
Therefore the statistically significant flux for these three red dwarfs is due to contamination from the nearby Crab pulsar, and cannot be attributed to these red dwarfs.

Similarly, for V1589 Cyg, we see a significant flux from 0.75-1.2 MeV (cf. Fig.~\ref{fig:RD5}). However, V1589 Cyg is within $3.7^{\circ}$ of Cygnus X-1, whose flux between 0.75-5 MeV is equal to $(9.78 \pm 0.92) \times 10^{-5} \text{ photons cm}^{-2} \text{ s}^{-1}$~\cite{Comptel00}. Therefore, we again considered a template of two point sources: one for  V1589 Cyg and another for Cygnus. With this change, once again we get only null detections at all energies and the SED plots  display  upper limits at all energies.




Therefore, we  conclude that  none of the eight TeV bright red dwarfs show non-zero flux in any energy bin between 0.75-30 MeV with significance $>3\sigma$.  We then calculate the 95\% confidence level upper limits for the differential photon flux (at 30 MeV), integral photon flux (from 0.75-30 MeV), and the integral energy flux.
These upper limits for all the three templates (along with two point source templates for the red dwarfs close to Crab and Cygnus) can be found in Table~\ref{tab:flux_limits}. The differential photon flux limits vary between $10^{-8}- 10^{-7} \rm{ph/cm^2/s/MeV}$, whereas the integral photon flux limits range between $10^{-6}-10^{-4}~\rm{ph/cm^2/s}$.  The integral energy flux limits range between $10^{-11}$ and $10^{-10}~\rm{erg/cm^2/s}$.

\renewcommand{\arraystretch}{1.25} 

\begin{table}[]
\caption{TS values for the Red Dwarfs for each of the red dwarfs using three different search templates, along with total exposure time, number of viewing periods and angular separation to the nearest source (within $4^{\circ}$) from the first COMPTEL source catalog~\cite{Comptel00}.}
\label{tab:Table_TS}
\begin{tabular}{|c|c|c|c|c|c|c|}
\hline
\textbf{\begin{tabular}[c]{@{}c@{}}Name of \\ Red Dwarf\end{tabular}} & \textbf{Template}        & \textbf{TS Value} & \textbf{\begin{tabular}[c]{@{}c@{}}Closest\\  COMPTEL source\\ within $4^{\circ}$\end{tabular}} & \textbf{\begin{tabular}[c]{@{}c@{}}Angular \\ separation \\ (deg)\end{tabular}} & \textbf{\begin{tabular}[c]{@{}c@{}}Number Of\\  Viewing Periods\end{tabular}} & \textbf{\begin{tabular}[c]{@{}c@{}}Exposure \\ ($\text{cm}^2$s)\end{tabular}} \\ \hline
\multirow{3}{*}{V388 Cas}                                             & Point Source             & 0.12              & \multirow{3}{*}{-}                                                                        & \multirow{3}{*}{-}                                                              & \multirow{3}{*}{10}                                                           & \multirow{3}{*}{$3.09 \times 10^{10}$}                                        \\ \cline{2-3}
                                                                      & Radial Disk              & 0.12              &                                                                                           &                                                                                 &                                                                               &                                                                       \\ \cline{2-3}
                                                                      & Radial Gaussian          & 0.12              &                                                                                           &                                                                                 &                                                                               &                                                                       \\ \hline
\multirow{3}{*}{V547 Cas}                                             & Point Source             & 0.2               & \multirow{3}{*}{-}                                                                        & \multirow{3}{*}{-}                                                              & \multirow{3}{*}{10}                                                           & \multirow{3}{*}{$3.09 \times 10^{10}$}                                        \\ \cline{2-3}
                                                                      & Radial Disk              & 0.2               &                                                                                           &                                                                                 &                                                                               &                                                                       \\ \cline{2-3}
                                                                      & Radial Gaussian          & 0.2               &                                                                                           &                                                                                 &                                                                               &                                                                       \\ \hline
\multirow{3}{*}{GJ 3684}                                              & Point Source             & $9.49 \times 10^{-3}$     & \multirow{3}{*}{-}                                                                        & \multirow{3}{*}{-}                                                              & \multirow{3}{*}{13}                                                           & \multirow{3}{*}{$3.66 \times 10^{10}$}                                        \\ \cline{2-3}
                                                                      & Radial Disk              & $9.34 \times 10^{-3}$     &                                                                                           &                                                                                 &                                                                               &                                                                       \\ \cline{2-3}
                                                                      & Radial Gaussian          & $9.29 \times 10^{-3}$     &                                                                                           &                                                                                 &                                                                               &                                                                       \\ \hline
\multirow{3}{*}{GL 851.1}                                             & Point Source             & $2.93 \times 10^{-2}$     & \multirow{3}{*}{-}                                                                        & \multirow{3}{*}{-}                                                              & \multirow{3}{*}{27}                                                           & \multirow{3}{*}{$6.47 \times 10^{10}$}                                        \\ \cline{2-3}
                                                                      & Radial Disk              & $2.96 \times 10^{-2}$     &                                                                                           &                                                                                 &                                                                               &                                                                       \\ \cline{2-3}
                                                                      & Radial Gaussian          & $2.97 \times 10^{-2}$     &                                                                                           &                                                                                 &                                                                               &                                                                       \\ \hline
\multirow{6}{*}{V780 Tau}                                             & Point Source             & 372.3            & \multirow{6}{*}{Crab pulsar}                                                              & \multirow{6}{*}{3.08}                                                           & \multirow{6}{*}{23}                                                           & \multirow{6}{*}{$6.55 \times 10^{10}$}                                        \\ \cline{2-3}
                                                                      & Radial Disk              & 372.7           &                                                                                           &                                                                                 &                                                                               &                                                                       \\ \cline{2-3}
                                                                      & Radial Gaussian          & 373.8           &                                                                                           &                                                                                 &                                                                               &                                                                       \\ \cline{2-3}
                                                                      & Point Source + Crab      & 0                 &                                                                                           &                                                                                 &                                                                               &                                                                       \\ \cline{2-3}
                                                                      & Radial Disk + Crab       & 0                 &                                                                                           &                                                                                 &                                                                               &                                                                       \\ \cline{2-3}
                                                                      & Radial Gaussian + Crab   & 0                 &                                                                                           &                                                                                 &                                                                               &                                                                       \\ \hline
\multirow{6}{*}{V962 Tau}                                             & Point Source             & 462.1           & \multirow{6}{*}{Crab Pulsar}                                                              & \multirow{6}{*}{2.75}                                                           & \multirow{6}{*}{24}                                                           & \multirow{6}{*}{$6.82 \times 10^{10}$}                                        \\ \cline{2-3}
                                                                      & Radial Disk              & 462.5          &                                                                                           &                                                                                 &                                                                               &                                                                       \\ \cline{2-3}
                                                                      & Radial Gaussian          & 463.7            &                                                                                           &                                                                                 &                                                                               &                                                                       \\ \cline{2-3}
                                                                      & Point Source + Crab      & 0                 &                                                                                           &                                                                                 &                                                                               &                                                                       \\ \cline{2-3}
                                                                      & Radial Disk + Crab       & 0                 &                                                                                           &                                                                                 &                                                                               &                                                                       \\ \cline{2-3}
                                                                      & Radial Gaussian + Crab   & 0                 &                                                                                           &                                                                                 &                                                                               &                                                                       \\ \hline
\multirow{6}{*}{GJ 1078}                                              & Point Source             & 497.2            & \multirow{6}{*}{Crab Pulsar}                                                              & \multirow{6}{*}{0.64}                                                           & \multirow{6}{*}{23}                                                           & \multirow{6}{*}{$6.55 \times 10^{10}$}                                        \\ \cline{2-3}
                                                                      & Radial Disk              & 497.6             &                                                                                           &                                                                                 &                                                                               &                                                                       \\ \cline{2-3}
                                                                      & Radial Gaussian          & 498.5            &                                                                                           &                                                                                 &                                                                               &                                                                       \\ \cline{2-3}
                                                                      & Point Source + Crab      & 0                 &                                                                                           &                                                                                 &                                                                               &                                                                       \\ \cline{2-3}
                                                                      & Radial Disk + Crab       & 0                 &                                                                                           &                                                                                 &                                                                               &                                                                       \\ \cline{2-3}
                                                                      & Radial Gaussian + Crab   & 0                 &                                                                                           &                                                                                 &                                                                               &                                                                       \\ \hline
\multirow{6}{*}{V1589 Cyg}                                            & Point Source             & 5.45              & \multirow{6}{*}{Cygnus X-1}                                                                   & \multirow{6}{*}{3.7}                                                            & \multirow{6}{*}{20}                                                           & \multirow{6}{*}{$3.88 \times 10^{10}$}                                        \\ \cline{2-3}
                                                                      & Radial Disk              & 5.46              &                                                                                           &                                                                                 &                                                                               &                                                                       \\ \cline{2-3}
                                                                      & Radial Gaussian          & 5.48              &                                                                                           &                                                                                 &                                                                               &                                                                       \\ \cline{2-3}
                                                                      & Point Source + Cygnus    & 0                 &                                                                                           &                                                                                 &                                                                               &                                                                       \\ \cline{2-3}
                                                                      & Radial Disk + Cygnus     & 0                 &                                                                                           &                                                                                 &                                                                               &                                                                       \\ \cline{2-3}
                                                                      & Radial Gaussian + Cygnus & 0                 &                                                                                           &                                                                                 &                                                                               &                                                                       \\ \hline
\end{tabular}
\end{table}

\renewcommand{\arraystretch}{1.25} 

\begin{table}[]
\caption{Results for  MeV gamma-ray emission from the eight red dwarfs searched using  COMPTEL data. We report $95\%$ c.l. upper limits for
differential flux (at 30 MeV), integral flux, and energy flux for all the three search templates used along with two source templates for the red dwarfs which reported significant detections at the lowest energies.}
\label{tab:flux_limits}
\begin{tabular}{|c|c|c|c|c|}
\hline
\textbf{\begin{tabular}[c]{@{}c@{}}Name of \\ Red Dwarf\end{tabular}} & \textbf{Template}                 & \textbf{\begin{tabular}[c]{@{}c@{}}Differential Flux \\ at 30 MeV\\  (ph/$\text{cm}^2$/s/MeV)\end{tabular}} & \textbf{\begin{tabular}[c]{@{}c@{}}Intergral Photon\\  Flux \\ (ph/$\text{cm}^2$/s)\end{tabular}} & \textbf{\begin{tabular}[c]{@{}c@{}}Integral Energy \\ Flux \\ (erg/$\text{cm}^2$/s)\end{tabular}} \\ \hline
\multirow{3}{*}{V388 Cas}                                    & Point Source             &$ < 3.57 \times 10^{-7}$                                                                             &$ < 9.83 \times 10^{-5}$                                                                   &$ < 1.75 \times 10^{-10}$                                                                  \\ \cline{2-5} 
                                                             & Radial Disk              &$ < 3.57 \times 10^{-7}$                                                                             &$ < 9.83 \times 10^{-5}$                                                                   &$ < 1.76 \times 10^{-10}$                                                                  \\ \cline{2-5} 
                                                             & Radial Gaussian          &$ < 3.56 \times 10^{-7}$                                                                             &$ < 9.85 \times 10^{-5}$                                                                   &$ < 1.76 \times 10^{-10}$                                                                  \\ \hline
\multirow{3}{*}{V547 Cas}                                    & Point Source             &$ < 3.32 \times 10^{-7}$                                                                             &$ < 1.14 \times 10^{-4}$                                                                   &$ < 1.35 \times 10^{-10}$                                                                  \\ \cline{2-5} 
                                                             & Radial Disk              &$ < 3.33 \times 10^{-7}$                                                                             &$ < 1.14 \times 10^{-4}$                                                                   &$ < 1.35 \times 10^{-10}$                                                                  \\ \cline{2-5} 
                                                             & Radial Gaussian          &$ < 3.34 \times 10^{-7}$                                                                             &$ < 1.14 \times 10^{-4}$                                                                   &$ < 1.36 \times 10^{-10}$                                                                  \\ \hline
\multirow{3}{*}{GJ 3684}                                     & Point Source             &$ < 9.31 \times 10^{-7}$                                                                             &$ < 5.73 \times 10^{-5}$                                                                   &$ < 1.77 \times 10^{-10}$                                                                  \\ \cline{2-5} 
                                                             & Radial Disk              &$ < 9.32 \times 10^{-7}$                                                                             &$ < 5.72 \times 10^{-5}$                                                                   &$ < 1.77 \times 10^{-10}$                                                                  \\ \cline{2-5} 
                                                             & Radial Gaussian          &$ < 9.36 \times 10^{-7}$                                                                             &$ < 5.73 \times 10^{-5}$                                                                   &$ < 1.77 \times 10^{-10}$                                                                  \\ \hline
\multirow{3}{*}{GL 851.1}                                    & Point Source             &$ < 3.34 \times 10^{-7}$                                                                             &$ < 5.54 \times 10^{-5}$                                                                   &$ < 1.52 \times 10^{-10}$                                                                  \\ \cline{2-5} 
                                                             & Radial Disk              &$ < 3.34 \times 10^{-7}$                                                                             &$ < 5.54 \times 10^{-5}$                                                                   &$ < 1.52 \times 10^{-10}$                                                                  \\ \cline{2-5} 
                                                             & Radial Gaussian          &$ < 3.35 \times 10^{-7}$                                                                             &$ < 5.56 \times 10^{-5}$                                                                   &$ < 1.53 \times 10^{-10}$                                                                  \\ \hline
\multirow{3}{*}{V780 Tau}                                    & Point Source + Crab      &$ < 1.02 \times 10^{-7}$                                                                             &$ < 5.29 \times 10^{-6}$                                                                   &$ < 1.73 \times 10^{-11}$                                                                  \\ \cline{2-5} 
                                                             & Radial Disk + Crab       &$ < 1.02 \times 10^{-7}$                                                                             &$ < 5.29 \times 10^{-6}$                                                                   &$ < 1.73 \times 10^{-11}$                                                                  \\ \cline{2-5} 
                                                             & Radial Gaussian + Crab   &$ < 1.02 \times 10^{-7}$                                                                             &$ < 5.30 \times 10^{-6}$                                                                   &$ < 1.73 \times 10^{-11}$                                                                  \\ \hline
\multirow{3}{*}{V962 Tau}                                    & Point Source + Crab      &$ < 8.40 \times 10^{-8}$                                                                             &$ < 4.62 \times 10^{-6}$                                                                   &$ < 1.50 \times 10^{-11}$                                                                  \\ \cline{2-5} 
                                                             & Radial Disk + Crab       &$ < 8.41 \times 10^{-8}$                                                                             &$ < 4.62 \times 10^{-6}$                                                                   &$ < 1.50 \times 10^{-11}$                                                                  \\ \cline{2-5} 
                                                             & Radial Gaussian + Crab   &$ < 8.43 \times 10^{-8}$                                                                             &$ < 4.63 \times 10^{-6}$                                                                   &$ < 1.50 \times 10^{-11}$                                                                  \\ \hline
\multirow{3}{*}{GJ 1078}                                     & Point Source + Crab      &$ < 7.21 \times 10^{-8}$                                                                             &$ < 4.60 \times 10^{-6}$                                                                   &$ < 1.40 \times 10^{-11}$                                                                  \\ \cline{2-5} 
                                                             & Radial Disk + Crab       &$ < 7.22 \times 10^{-8}$                                                                             &$ < 4.60 \times 10^{-6}$                                                                   &$ < 1.4 \times 10^{-11}$                                                                   \\ \cline{2-5} 
                                                             & Radial Gaussian + Crab   &$ < 7.23 \times 10^{-8}$                                                                             &$ < 4.62 \times 10^{-6}$                                                                   &$ < 1.40 \times 10^{-11}$                                                                  \\ \hline
\multirow{3}{*}{V1589 Cyg}                                  & Point Source + Cygnus    &$ < 8.64 \times 10^{-8}$                                                                             &$ < 4.86 \times 10^{-6 }$                                                                  &$ < 1.62 \times 10^{-11}$                                                                  \\ \cline{2-5} 
                                                             & Radial Disk + Cygnus     &$ < 8.63 \times 10^{-8}$                                                                             &$ < 4.86 \times 10^{-6}$                                                                   &$ < 1.62 \times 10^{-11}$                                                                  \\ \cline{2-5} 
                                                             & Radial Gaussian + Cygnus &$ < 8.63 \times 10^{-8}$                                                                             &$ < 4.86 \times 10^{-6}$                                                                   &$ < 1.62 \times 10^{-11}$                                                                  \\ \hline
\end{tabular}
\end{table}

\section{Implications of our results}
\label{sec:implications}
We now briefly discuss  some of the  implications of our flux limits in light of the SHALON results and theoretical models. The SHALON telescope detected VHE gamma-ray emission from 1-20 TeV with a spectral index between 2.6 and 3.2~\cite{Shalon}. The differential energy spectrum for   V388 Cas is given by $\frac{dN}{dE} \approx 0.9 \times 10^{-12} \left[\rm{E}/ (1~\rm{TeV})\right]^{-2.5} \rm{TeV^{-1} cm^{-2} s^{-1}}$. All other red dwarfs have similar spectral indices. \rthis{By extrapolating this spectral index  to the COMPTEL energy range, the differential photon flux at 10 MeV} is $E^2 \frac{dN}{dE} \approx 4  \times  10^{-10} \rm{ergs/cm^2/s}$ at 10 MeV. For V388 Cas, we obtain a 95\% upper limit of $10^{-10} \rm{ergs/cm^2/s}$, which is around the same order of magnitude as the extrapolated energy flux. Therefore prima-facie, our upper limits on the energy  flux are of the same order of magnitude as expected  from a  naive  extrapolation of the observed TeV spectrum to MeV energies. Since all other red dwarfs show similar spectral indices and differential flux, we expect this conclusion to hold true for all red dwarfs. 
We also note that the SHALON TeV emission  was detected during the flaring states of the   red dwarfs, whereas for our analysis we have considered the full COMPTEL dataset, where these red dwarfs could have been in quiescent state. Therefore, the expected \rthis{MeV} flux  during the quiescent phase would be expected to be much  smaller than that obtained during the flaring phase. \rthis{Note that one could also potentially look for transient MeV emission from red dwarfs by also looking   for temporal coincidences with  flares at other wavelengths.}

Some of the proposed  mechanisms  for gamma-ray emission from red dwarfs include  magnetic reconnection or CME-driven shock waves~\cite{Cliver,Ohm}  and  spallation from  interactions of background 
cosmic rays with the the red dwarf atmospheres, similar to that proposed for  the Sun~\cite{Zhou}. The SED for the CME driven shock wave scenario has been calculated in ~\cite{Ohm} for nearby M-dwarfs based on the energetics of DG CVn superflare and extrapolations from solar flares. This work has estimated the energy flux to be between $10^{-13}$  and $2 \times 10^{-12} \rm{ergs ~cm^{-2}~s^{-1}}$ at energies between 10 and 30 MeV (cf. Fig. 1 of ~\cite{Ohm}). This is  one to two orders of magnitude lower than our limits. Therefore, we do not rule \rthis{out} this model.
The expected energy flux  due to interactions from background cosmic rays  has been estimated for V962 Tau at GeV energies and found to be $10^{-29} \rm{ergs/cm^2/s}$~\cite{Zhou}. Therefore, the expected flux from this mechanism is too small to be detected in COMPTEL.

\section{\label{sec:conclusions}Conclusions}\protect
Recently, very high energy gamma-ray emission has been detected from eight red dwarfs between 800 GeV to 20 TeV, namely V388~Cas,  V547~Cas, V780~Tau, V962~Tau, V1589~Cyg, GJ~1078,
GJ~3684, and GL~851.1 from the SHALON atmospheric Cherenkov telescope~\cite{Shalon}. A recent work also  looked for GeV gamma-ray emission between 0.2-500 GeV with Fermi-LAT~\cite{Huang24}. Meanwhile, the  soft gamma-ray energy band  (between $\sim$ 1-30 MeV) is unchartered territory and has not been imaged by any other gamma-ray telescope after the decommissioning of the COMPTEL  MeV telescope in 2000.
About two years ago, new software has been released to facilitate seamless analysis  of the archival COMPTEL data~\cite{Knod}.

Motivated by all these considerations,  we search for soft MeV gamma-ray emission (0.75 - 30 MeV) from these aforementioned eight red dwarfs using archival data from the COMPTEL gamma-ray telescope. For our analysis, we used three search templates: point source, radial disk, and radial Gaussian (similar to our recent work on searching from MeV emission from galaxy clusters~\cite{Manna}). 

Four of the red dwarfs  V780 Tau, V962 Tau, V1589 Cyg, and  GJ 1078 
are within the proximity of the Crab pulsar or Cygnus. Once we account for the contamination from Crab and Cygnus,   we obtain null detections for the aforementioned three red dwarfs at all energies. Two other red dwarfs,  GJ 3684 and GL 851.1 also show null detections at all energies. Finally the remaining two red dwarfs (V388 Cas and V547 Cas) report a non-zero flux  in only one energy bin at 13.39 MeV and 1.06 MeV, respectively. However the statistical significance of this is signal is only 0.59$\sigma$ and 1.74$\sigma$, respectively and therefore can  only be considered as  a fluctuation. The SED for all these red dwarfs can be found in Figs.~\ref{fig:RD1} to Figs~\ref{fig:RD5}.

Therefore, we do not detect any signal from these red dwarfs at soft MeV energies with significance $> 3\sigma$ using archival data from the COMPTEL gamma-ray telescope. We then report the 95\% c.l.  upper limits on the integral and differential photon flux (at 30 MeV), along with integrated energy flux limits. These limits have been collated in Table~\ref{tab:flux_limits}.  The integral energy flux limits we  obtain range between $10^{-11}-10^{-10} \rm{ergs/cm^2/s}$.

Our observed upper limits are roughly of the same order of magnitude  as the expected flux at MeV energies, estimated from extrapolating the TeV spectrum. 
The expected energy flux in the MeV energy for one theoretical model involving CME-driven shock waves~\cite{Ohm} is about one to two orders of magnitude smaller than our upper limits, implying that COMPTEL is not sensitive enough to detect MeV gamma-ray emission expected from this model.  A large number of    gamma-ray missions such as ASTROMEV~\cite{MeVmissions}, COSI~\cite{COSI}, GECCO~\cite{GECCO}, GAMMA-400~\cite{Gamma400}, Advanced Compton Telescope~\cite{Boggs2006}, AdEPT~\cite{Hunter2014}, PANGU~\cite{Wu2014}, GRAMS~\cite{Grams}, MAST~\cite{MAST}, etc have either been proposed or scheduled to be launched in the next two decades. These forthcoming missions  should be much more sensitive than COMPTEL and hence could potentially detect a signal from these red dwarfs at MeV energies.

\begin{acknowledgments}
NS has been supported by a Summer Undergraduate Research Exposure (SURE) Internship at IIT Hyderabad. 
SM is supported by  Government of India, Ministry of Education (MOE) fellowship.  We are also very grateful to 
J\"urgen Kn\"odlseder for  thoroughly   explaining to us the  nuances of the COMPTEL analysis software and generously  helping us in answering all our doubts. Finally, we acknowledge the anonymous referee for very useful feedback on our manuscript.

\end{acknowledgments}

\bibliography{references}

\begin{thebibliography}{35}
\expandafter\ifx\csname natexlab\endcsname\relax\def\natexlab#1{#1}\fi
\expandafter\ifx\csname bibnamefont\endcsname\relax
  \def\bibnamefont#1{#1}\fi
\expandafter\ifx\csname bibfnamefont\endcsname\relax
  \def\bibfnamefont#1{#1}\fi
\expandafter\ifx\csname citenamefont\endcsname\relax
  \def\citenamefont#1{#1}\fi
\expandafter\ifx\csname url\endcsname\relax
  \def\url#1{\texttt{#1}}\fi
\expandafter\ifx\csname urlprefix\endcsname\relax\def\urlprefix{URL }\fi
\providecommand{\bibinfo}[2]{#2}
\providecommand{\eprint}[2][]{\url{#2}}

\bibitem[{\citenamefont{{Edgeworth}}(1946)}]{Edgeworth}
\bibinfo{author}{\bibfnamefont{K.~E.} \bibnamefont{{Edgeworth}}}, \bibinfo{journal}{\nat} \textbf{\bibinfo{volume}{157}}, \bibinfo{pages}{481} (\bibinfo{year}{1946}).

\bibitem[{\citenamefont{{Aschwanden} and {G{\"u}del}}(2021)}]{Aschwanden}
\bibinfo{author}{\bibfnamefont{M.~J.} \bibnamefont{{Aschwanden}}} \bibnamefont{and} \bibinfo{author}{\bibfnamefont{M.}~\bibnamefont{{G{\"u}del}}}, \bibinfo{journal}{\apj} \textbf{\bibinfo{volume}{910}}, \bibinfo{eid}{41} (\bibinfo{year}{2021}), \eprint{2106.06490}.

\bibitem[{\citenamefont{{Yang} et~al.}(2017)\citenamefont{{Yang}, {Liu}, {Gao}, {Fang}, {Guo}, {Zhang}, {Hou}, {Wang}, and {Cao}}}]{Yang17}
\bibinfo{author}{\bibfnamefont{H.}~\bibnamefont{{Yang}}}, \bibinfo{author}{\bibfnamefont{J.}~\bibnamefont{{Liu}}}, \bibinfo{author}{\bibfnamefont{Q.}~\bibnamefont{{Gao}}}, \bibinfo{author}{\bibfnamefont{X.}~\bibnamefont{{Fang}}}, \bibinfo{author}{\bibfnamefont{J.}~\bibnamefont{{Guo}}}, \bibinfo{author}{\bibfnamefont{Y.}~\bibnamefont{{Zhang}}}, \bibinfo{author}{\bibfnamefont{Y.}~\bibnamefont{{Hou}}}, \bibinfo{author}{\bibfnamefont{Y.}~\bibnamefont{{Wang}}}, \bibnamefont{and} \bibinfo{author}{\bibfnamefont{Z.}~\bibnamefont{{Cao}}}, \bibinfo{journal}{\apj} \textbf{\bibinfo{volume}{849}}, \bibinfo{eid}{36} (\bibinfo{year}{2017}).

\bibitem[{\citenamefont{{Drake} et~al.}(2014)\citenamefont{{Drake}, {Osten}, {Page}, {Kennea}, {Oates}, {Krimm}, and {Gehrels}}}]{Drake}
\bibinfo{author}{\bibfnamefont{S.}~\bibnamefont{{Drake}}}, \bibinfo{author}{\bibfnamefont{R.}~\bibnamefont{{Osten}}}, \bibinfo{author}{\bibfnamefont{K.~L.} \bibnamefont{{Page}}}, \bibinfo{author}{\bibfnamefont{J.~A.} \bibnamefont{{Kennea}}}, \bibinfo{author}{\bibfnamefont{S.~R.} \bibnamefont{{Oates}}}, \bibinfo{author}{\bibfnamefont{H.}~\bibnamefont{{Krimm}}}, \bibnamefont{and} \bibinfo{author}{\bibfnamefont{N.}~\bibnamefont{{Gehrels}}}, \bibinfo{journal}{The Astronomer's Telegram} \textbf{\bibinfo{volume}{6121}}, \bibinfo{pages}{1} (\bibinfo{year}{2014}).

\bibitem[{\citenamefont{{Fender} et~al.}(2015)\citenamefont{{Fender}, {Anderson}, {Osten}, {Staley}, {Rumsey}, {Grainge}, and {Saunders}}}]{Fender}
\bibinfo{author}{\bibfnamefont{R.~P.} \bibnamefont{{Fender}}}, \bibinfo{author}{\bibfnamefont{G.~E.} \bibnamefont{{Anderson}}}, \bibinfo{author}{\bibfnamefont{R.}~\bibnamefont{{Osten}}}, \bibinfo{author}{\bibfnamefont{T.}~\bibnamefont{{Staley}}}, \bibinfo{author}{\bibfnamefont{C.}~\bibnamefont{{Rumsey}}}, \bibinfo{author}{\bibfnamefont{K.}~\bibnamefont{{Grainge}}}, \bibnamefont{and} \bibinfo{author}{\bibfnamefont{R.~D.~E.} \bibnamefont{{Saunders}}}, \bibinfo{journal}{\mnras} \textbf{\bibinfo{volume}{446}}, \bibinfo{pages}{L66} (\bibinfo{year}{2015}), \eprint{1410.1545}.

\bibitem[{\citenamefont{{Loh} et~al.}(2017)\citenamefont{{Loh}, {Corbel}, and {Dubus}}}]{Loh}
\bibinfo{author}{\bibfnamefont{A.}~\bibnamefont{{Loh}}}, \bibinfo{author}{\bibfnamefont{S.}~\bibnamefont{{Corbel}}}, \bibnamefont{and} \bibinfo{author}{\bibfnamefont{G.}~\bibnamefont{{Dubus}}}, \bibinfo{journal}{\mnras} \textbf{\bibinfo{volume}{467}}, \bibinfo{pages}{4462} (\bibinfo{year}{2017}), \eprint{1702.03754}.

\bibitem[{\citenamefont{{Song} and {Paglione}}(2020)}]{Song}
\bibinfo{author}{\bibfnamefont{Y.}~\bibnamefont{{Song}}} \bibnamefont{and} \bibinfo{author}{\bibfnamefont{T.~A.~D.} \bibnamefont{{Paglione}}}, \bibinfo{journal}{\apj} \textbf{\bibinfo{volume}{900}}, \bibinfo{eid}{185} (\bibinfo{year}{2020}), \eprint{2008.01143}.

\bibitem[{\citenamefont{{Sinitsyna} et~al.}(2019)\citenamefont{{Sinitsyna}, {Sinitsyna}, and {Stozhkov}}}]{Shalon}
\bibinfo{author}{\bibfnamefont{V.~G.} \bibnamefont{{Sinitsyna}}}, \bibinfo{author}{\bibfnamefont{V.~Y.} \bibnamefont{{Sinitsyna}}}, \bibnamefont{and} \bibinfo{author}{\bibfnamefont{Y.~I.} \bibnamefont{{Stozhkov}}}, \bibinfo{journal}{Advances in Space Research} \textbf{\bibinfo{volume}{64}}, \bibinfo{pages}{2585} (\bibinfo{year}{2019}).

\bibitem[{\citenamefont{{Gershberg} et~al.}(1999)\citenamefont{{Gershberg}, {Katsova}, {Lovkaya}, {Terebizh}, and {Shakhovskaya}}}]{Gershberg}
\bibinfo{author}{\bibfnamefont{R.~E.} \bibnamefont{{Gershberg}}}, \bibinfo{author}{\bibfnamefont{M.~M.} \bibnamefont{{Katsova}}}, \bibinfo{author}{\bibfnamefont{M.~N.} \bibnamefont{{Lovkaya}}}, \bibinfo{author}{\bibfnamefont{A.~V.} \bibnamefont{{Terebizh}}}, \bibnamefont{and} \bibinfo{author}{\bibfnamefont{N.~I.} \bibnamefont{{Shakhovskaya}}}, \bibinfo{journal}{\aaps} \textbf{\bibinfo{volume}{139}}, \bibinfo{pages}{555} (\bibinfo{year}{1999}).

\bibitem[{\citenamefont{{Sinitsyna} et~al.}(2022)\citenamefont{{Sinitsyna}, {Sinitsyna}, and {Stozhkov}}}]{rdcosmicrays}
\bibinfo{author}{\bibfnamefont{V.~Y.} \bibnamefont{{Sinitsyna}}}, \bibinfo{author}{\bibfnamefont{V.~G.} \bibnamefont{{Sinitsyna}}}, \bibnamefont{and} \bibinfo{author}{\bibfnamefont{Y.~I.} \bibnamefont{{Stozhkov}}}, in \emph{\bibinfo{booktitle}{European Physical Journal Web of Conferences}} (\bibinfo{year}{2022}), vol. \bibinfo{volume}{260} of \emph{\bibinfo{series}{European Physical Journal Web of Conferences}}, p. \bibinfo{pages}{11033}.

\bibitem[{\citenamefont{{Huang} et~al.}(2024)\citenamefont{{Huang}, {Zhang}, {Chen}, and {Zhong}}}]{Huang24}
\bibinfo{author}{\bibfnamefont{C.}~\bibnamefont{{Huang}}}, \bibinfo{author}{\bibfnamefont{X.}~\bibnamefont{{Zhang}}}, \bibinfo{author}{\bibfnamefont{Y.}~\bibnamefont{{Chen}}}, \bibnamefont{and} \bibinfo{author}{\bibfnamefont{W.}~\bibnamefont{{Zhong}}}, \bibinfo{journal}{\apj} \textbf{\bibinfo{volume}{965}}, \bibinfo{eid}{26} (\bibinfo{year}{2024}), \eprint{2403.12524}.

\bibitem[{\citenamefont{{Manna} and {Desai}}(2024{\natexlab{a}})}]{Manna}
\bibinfo{author}{\bibfnamefont{S.}~\bibnamefont{{Manna}}} \bibnamefont{and} \bibinfo{author}{\bibfnamefont{S.}~\bibnamefont{{Desai}}}, \bibinfo{journal}{\jcap} \textbf{\bibinfo{volume}{2024}}, \bibinfo{eid}{013} (\bibinfo{year}{2024}{\natexlab{a}}), \eprint{2401.13240}.

\bibitem[{\citenamefont{{Schoenfelder} et~al.}(1993)\citenamefont{{Schoenfelder}, {Aarts}, {Bennett}, {de Boer}, {Clear}, {Collmar}, {Connors}, {Deerenberg}, {Diehl}, {von Dordrecht} et~al.}}]{Comptel93}
\bibinfo{author}{\bibfnamefont{V.}~\bibnamefont{{Schoenfelder}}}, \bibinfo{author}{\bibfnamefont{H.}~\bibnamefont{{Aarts}}}, \bibinfo{author}{\bibfnamefont{K.}~\bibnamefont{{Bennett}}}, \bibinfo{author}{\bibfnamefont{H.}~\bibnamefont{{de Boer}}}, \bibinfo{author}{\bibfnamefont{J.}~\bibnamefont{{Clear}}}, \bibinfo{author}{\bibfnamefont{W.}~\bibnamefont{{Collmar}}}, \bibinfo{author}{\bibfnamefont{A.}~\bibnamefont{{Connors}}}, \bibinfo{author}{\bibfnamefont{A.}~\bibnamefont{{Deerenberg}}}, \bibinfo{author}{\bibfnamefont{R.}~\bibnamefont{{Diehl}}}, \bibinfo{author}{\bibfnamefont{A.}~\bibnamefont{{von Dordrecht}}}, \bibnamefont{et~al.}, \bibinfo{journal}{\apjs} \textbf{\bibinfo{volume}{86}}, \bibinfo{pages}{657} (\bibinfo{year}{1993}).

\bibitem[{\citenamefont{{Strong} and {Collmar}}(2019)}]{Strong}
\bibinfo{author}{\bibfnamefont{A.}~\bibnamefont{{Strong}}} \bibnamefont{and} \bibinfo{author}{\bibfnamefont{W.}~\bibnamefont{{Collmar}}}, \bibinfo{journal}{\memsai} \textbf{\bibinfo{volume}{90}}, \bibinfo{pages}{297} (\bibinfo{year}{2019}), \eprint{1907.07454}.

\bibitem[{\citenamefont{{Sch{\"o}nfelder} et~al.}(2000)\citenamefont{{Sch{\"o}nfelder}, {Bennett}, {Blom}, {Bloemen}, {Collmar}, {Connors}, {Diehl}, {Hermsen}, {Iyudin}, {Kippen} et~al.}}]{Comptel00}
\bibinfo{author}{\bibfnamefont{V.}~\bibnamefont{{Sch{\"o}nfelder}}}, \bibinfo{author}{\bibfnamefont{K.}~\bibnamefont{{Bennett}}}, \bibinfo{author}{\bibfnamefont{J.~J.} \bibnamefont{{Blom}}}, \bibinfo{author}{\bibfnamefont{H.}~\bibnamefont{{Bloemen}}}, \bibinfo{author}{\bibfnamefont{W.}~\bibnamefont{{Collmar}}}, \bibinfo{author}{\bibfnamefont{A.}~\bibnamefont{{Connors}}}, \bibinfo{author}{\bibfnamefont{R.}~\bibnamefont{{Diehl}}}, \bibinfo{author}{\bibfnamefont{W.}~\bibnamefont{{Hermsen}}}, \bibinfo{author}{\bibfnamefont{A.}~\bibnamefont{{Iyudin}}}, \bibinfo{author}{\bibfnamefont{R.~M.} \bibnamefont{{Kippen}}}, \bibnamefont{et~al.}, \bibinfo{journal}{\aaps} \textbf{\bibinfo{volume}{143}}, \bibinfo{pages}{145} (\bibinfo{year}{2000}), \eprint{astro-ph/0002366}.

\bibitem[{\citenamefont{{Kn{\"o}dlseder} et~al.}(2016)\citenamefont{{Kn{\"o}dlseder}, {Mayer}, {Deil}, {Cayrou}, {Owen}, {Kelley-Hoskins}, {Lu}, {Buehler}, {Forest}, {Louge} et~al.}}]{ctools}
\bibinfo{author}{\bibfnamefont{J.}~\bibnamefont{{Kn{\"o}dlseder}}}, \bibinfo{author}{\bibfnamefont{M.}~\bibnamefont{{Mayer}}}, \bibinfo{author}{\bibfnamefont{C.}~\bibnamefont{{Deil}}}, \bibinfo{author}{\bibfnamefont{J.~B.} \bibnamefont{{Cayrou}}}, \bibinfo{author}{\bibfnamefont{E.}~\bibnamefont{{Owen}}}, \bibinfo{author}{\bibfnamefont{N.}~\bibnamefont{{Kelley-Hoskins}}}, \bibinfo{author}{\bibfnamefont{C.~C.} \bibnamefont{{Lu}}}, \bibinfo{author}{\bibfnamefont{R.}~\bibnamefont{{Buehler}}}, \bibinfo{author}{\bibfnamefont{F.}~\bibnamefont{{Forest}}}, \bibinfo{author}{\bibfnamefont{T.}~\bibnamefont{{Louge}}}, \bibnamefont{et~al.}, \bibinfo{journal}{\aap} \textbf{\bibinfo{volume}{593}}, \bibinfo{eid}{A1} (\bibinfo{year}{2016}), \eprint{1606.00393}.

\bibitem[{\citenamefont{{Kn{\"o}dlseder} et~al.}(2022)\citenamefont{{Kn{\"o}dlseder}, {Collmar}, {Jarry}, and {McConnell}}}]{Knod}
\bibinfo{author}{\bibfnamefont{J.}~\bibnamefont{{Kn{\"o}dlseder}}}, \bibinfo{author}{\bibfnamefont{W.}~\bibnamefont{{Collmar}}}, \bibinfo{author}{\bibfnamefont{M.}~\bibnamefont{{Jarry}}}, \bibnamefont{and} \bibinfo{author}{\bibfnamefont{M.}~\bibnamefont{{McConnell}}}, \bibinfo{journal}{\aap} \textbf{\bibinfo{volume}{665}}, \bibinfo{eid}{A84} (\bibinfo{year}{2022}), \eprint{2207.13404}.

\bibitem[{\citenamefont{{Wood} et~al.}(2017)\citenamefont{{Wood}, {Caputo}, {Charles}, {Di Mauro}, {Magill}, {Perkins}, and {Fermi-LAT Collaboration}}}]{Wood2017}
\bibinfo{author}{\bibfnamefont{M.}~\bibnamefont{{Wood}}}, \bibinfo{author}{\bibfnamefont{R.}~\bibnamefont{{Caputo}}}, \bibinfo{author}{\bibfnamefont{E.}~\bibnamefont{{Charles}}}, \bibinfo{author}{\bibfnamefont{M.}~\bibnamefont{{Di Mauro}}}, \bibinfo{author}{\bibfnamefont{J.}~\bibnamefont{{Magill}}}, \bibinfo{author}{\bibfnamefont{J.~S.} \bibnamefont{{Perkins}}}, \bibnamefont{and} \bibinfo{author}{\bibnamefont{{Fermi-LAT Collaboration}}}, in \emph{\bibinfo{booktitle}{35th International Cosmic Ray Conference (ICRC2017)}} (\bibinfo{year}{2017}), vol. \bibinfo{volume}{301} of \emph{\bibinfo{series}{International Cosmic Ray Conference}}, p. \bibinfo{pages}{824}, \eprint{1707.09551}.

\bibitem[{\citenamefont{{Mattox} et~al.}(1996)\citenamefont{{Mattox}, {Bertsch}, {Chiang}, {Dingus}, {Digel}, {Esposito}, {Fierro}, {Hartman}, {Hunter}, {Kanbach} et~al.}}]{Mattox1996}
\bibinfo{author}{\bibfnamefont{J.~R.} \bibnamefont{{Mattox}}}, \bibinfo{author}{\bibfnamefont{D.~L.} \bibnamefont{{Bertsch}}}, \bibinfo{author}{\bibfnamefont{J.}~\bibnamefont{{Chiang}}}, \bibinfo{author}{\bibfnamefont{B.~L.} \bibnamefont{{Dingus}}}, \bibinfo{author}{\bibfnamefont{S.~W.} \bibnamefont{{Digel}}}, \bibinfo{author}{\bibfnamefont{J.~A.} \bibnamefont{{Esposito}}}, \bibinfo{author}{\bibfnamefont{J.~M.} \bibnamefont{{Fierro}}}, \bibinfo{author}{\bibfnamefont{R.~C.} \bibnamefont{{Hartman}}}, \bibinfo{author}{\bibfnamefont{S.~D.} \bibnamefont{{Hunter}}}, \bibinfo{author}{\bibfnamefont{G.}~\bibnamefont{{Kanbach}}}, \bibnamefont{et~al.}, \bibinfo{journal}{\apj} \textbf{\bibinfo{volume}{461}}, \bibinfo{pages}{396} (\bibinfo{year}{1996}).

\bibitem[{\citenamefont{Wilks}(1938)}]{Wilks}
\bibinfo{author}{\bibfnamefont{S.~S.} \bibnamefont{Wilks}}, \bibinfo{journal}{The annals of mathematical statistics} \textbf{\bibinfo{volume}{9}}, \bibinfo{pages}{60} (\bibinfo{year}{1938}).

\bibitem[{\citenamefont{{Manna} and {Desai}}(2024{\natexlab{b}})}]{MannaFermi}
\bibinfo{author}{\bibfnamefont{S.}~\bibnamefont{{Manna}}} \bibnamefont{and} \bibinfo{author}{\bibfnamefont{S.}~\bibnamefont{{Desai}}}, \bibinfo{journal}{\jcap} \textbf{\bibinfo{volume}{2024}}, \bibinfo{eid}{017} (\bibinfo{year}{2024}{\natexlab{b}}), \eprint{2310.07519}.

\bibitem[{\citenamefont{{Pasumarti} and {Desai}}(2022)}]{Pasumarti}
\bibinfo{author}{\bibfnamefont{V.}~\bibnamefont{{Pasumarti}}} \bibnamefont{and} \bibinfo{author}{\bibfnamefont{S.}~\bibnamefont{{Desai}}}, \bibinfo{journal}{\jcap} \textbf{\bibinfo{volume}{2022}}, \bibinfo{eid}{002} (\bibinfo{year}{2022}), \eprint{2210.12804}.

\bibitem[{\citenamefont{{Pasumarti} and {Desai}}(2024)}]{Pasumarti2}
\bibinfo{author}{\bibfnamefont{V.}~\bibnamefont{{Pasumarti}}} \bibnamefont{and} \bibinfo{author}{\bibfnamefont{S.}~\bibnamefont{{Desai}}}, \bibinfo{journal}{\jcap} \textbf{\bibinfo{volume}{2024}}, \bibinfo{eid}{010} (\bibinfo{year}{2024}), \eprint{2306.03427}.

\bibitem[{\citenamefont{{Cliver} et~al.}(2022)\citenamefont{{Cliver}, {Schrijver}, {Shibata}, and {Usoskin}}}]{Cliver}
\bibinfo{author}{\bibfnamefont{E.~W.} \bibnamefont{{Cliver}}}, \bibinfo{author}{\bibfnamefont{C.~J.} \bibnamefont{{Schrijver}}}, \bibinfo{author}{\bibfnamefont{K.}~\bibnamefont{{Shibata}}}, \bibnamefont{and} \bibinfo{author}{\bibfnamefont{I.~G.} \bibnamefont{{Usoskin}}}, \bibinfo{journal}{Living Reviews in Solar Physics} \textbf{\bibinfo{volume}{19}}, \bibinfo{eid}{2} (\bibinfo{year}{2022}), \eprint{2205.09265}.

\bibitem[{\citenamefont{{Ohm} and {Hoischen}}(2018)}]{Ohm}
\bibinfo{author}{\bibfnamefont{S.}~\bibnamefont{{Ohm}}} \bibnamefont{and} \bibinfo{author}{\bibfnamefont{C.}~\bibnamefont{{Hoischen}}}, \bibinfo{journal}{\mnras} \textbf{\bibinfo{volume}{474}}, \bibinfo{pages}{1335} (\bibinfo{year}{2018}), \eprint{1710.09385}.

\bibitem[{\citenamefont{{Zhou} et~al.}(2017)\citenamefont{{Zhou}, {Ng}, {Beacom}, and {Peter}}}]{Zhou}
\bibinfo{author}{\bibfnamefont{B.}~\bibnamefont{{Zhou}}}, \bibinfo{author}{\bibfnamefont{K.~C.~Y.} \bibnamefont{{Ng}}}, \bibinfo{author}{\bibfnamefont{J.~F.} \bibnamefont{{Beacom}}}, \bibnamefont{and} \bibinfo{author}{\bibfnamefont{A.~H.~G.} \bibnamefont{{Peter}}}, \bibinfo{journal}{\prd} \textbf{\bibinfo{volume}{96}}, \bibinfo{eid}{023015} (\bibinfo{year}{2017}), \eprint{1612.02420}.

\bibitem[{\citenamefont{{De Angelis} et~al.}(2021)\citenamefont{{De Angelis}, {Tatischeff}, {Argan}, {Brandt}, {Bulgarelli}, {Bykov}, {Costantini}, {Curado da Silva}, {Grenier}, {Hanlon} et~al.}}]{MeVmissions}
\bibinfo{author}{\bibfnamefont{A.}~\bibnamefont{{De Angelis}}}, \bibinfo{author}{\bibfnamefont{V.}~\bibnamefont{{Tatischeff}}}, \bibinfo{author}{\bibfnamefont{A.}~\bibnamefont{{Argan}}}, \bibinfo{author}{\bibfnamefont{S.}~\bibnamefont{{Brandt}}}, \bibinfo{author}{\bibfnamefont{A.}~\bibnamefont{{Bulgarelli}}}, \bibinfo{author}{\bibfnamefont{A.}~\bibnamefont{{Bykov}}}, \bibinfo{author}{\bibfnamefont{E.}~\bibnamefont{{Costantini}}}, \bibinfo{author}{\bibfnamefont{R.}~\bibnamefont{{Curado da Silva}}}, \bibinfo{author}{\bibfnamefont{I.~A.} \bibnamefont{{Grenier}}}, \bibinfo{author}{\bibfnamefont{L.}~\bibnamefont{{Hanlon}}}, \bibnamefont{et~al.}, \bibinfo{journal}{Experimental Astronomy} \textbf{\bibinfo{volume}{51}}, \bibinfo{pages}{1225} (\bibinfo{year}{2021}), \eprint{2102.02460}.

\bibitem[{\citenamefont{{Zoglauer} et~al.}(2021)\citenamefont{{Zoglauer}, {Siegert}, {Lowell}, {Mochizuki}, {Kierans}, {Sleator}, {Hartmann}, {Lazar}, {Gulick}, {Beechert} et~al.}}]{COSI}
\bibinfo{author}{\bibfnamefont{A.}~\bibnamefont{{Zoglauer}}}, \bibinfo{author}{\bibfnamefont{T.}~\bibnamefont{{Siegert}}}, \bibinfo{author}{\bibfnamefont{A.}~\bibnamefont{{Lowell}}}, \bibinfo{author}{\bibfnamefont{B.}~\bibnamefont{{Mochizuki}}}, \bibinfo{author}{\bibfnamefont{C.}~\bibnamefont{{Kierans}}}, \bibinfo{author}{\bibfnamefont{C.}~\bibnamefont{{Sleator}}}, \bibinfo{author}{\bibfnamefont{D.~H.} \bibnamefont{{Hartmann}}}, \bibinfo{author}{\bibfnamefont{H.}~\bibnamefont{{Lazar}}}, \bibinfo{author}{\bibfnamefont{H.}~\bibnamefont{{Gulick}}}, \bibinfo{author}{\bibfnamefont{J.}~\bibnamefont{{Beechert}}}, \bibnamefont{et~al.}, \bibinfo{journal}{arXiv e-prints} \bibinfo{eid}{arXiv:2102.13158} (\bibinfo{year}{2021}), \eprint{2102.13158}.

\bibitem[{\citenamefont{{Orlando} et~al.}(2022)\citenamefont{{Orlando}, {Bottacini}, {Moiseev}, {Bodaghee}, {Collmar}, {Ensslin}, {Moskalenko}, {Negro}, {Profumo}, {Digel} et~al.}}]{GECCO}
\bibinfo{author}{\bibfnamefont{E.}~\bibnamefont{{Orlando}}}, \bibinfo{author}{\bibfnamefont{E.}~\bibnamefont{{Bottacini}}}, \bibinfo{author}{\bibfnamefont{A.~A.} \bibnamefont{{Moiseev}}}, \bibinfo{author}{\bibfnamefont{A.}~\bibnamefont{{Bodaghee}}}, \bibinfo{author}{\bibfnamefont{W.}~\bibnamefont{{Collmar}}}, \bibinfo{author}{\bibfnamefont{T.}~\bibnamefont{{Ensslin}}}, \bibinfo{author}{\bibfnamefont{I.~V.} \bibnamefont{{Moskalenko}}}, \bibinfo{author}{\bibfnamefont{M.}~\bibnamefont{{Negro}}}, \bibinfo{author}{\bibfnamefont{S.}~\bibnamefont{{Profumo}}}, \bibinfo{author}{\bibfnamefont{S.~W.} \bibnamefont{{Digel}}}, \bibnamefont{et~al.}, \bibinfo{journal}{\jcap} \textbf{\bibinfo{volume}{2022}}, \bibinfo{eid}{036} (\bibinfo{year}{2022}), \eprint{2112.07190}.

\bibitem[{\citenamefont{{Galper} et~al.}(2014)\citenamefont{{Galper}, {Bonvicini}, {Topchiev}, {Adriani}, {Aptekar}, {Arkhangelskaja}, {Arkhangelskiy}, {Bergstrom}, {Berti}, {Bigongiari} et~al.}}]{Gamma400}
\bibinfo{author}{\bibfnamefont{A.~M.} \bibnamefont{{Galper}}}, \bibinfo{author}{\bibfnamefont{V.}~\bibnamefont{{Bonvicini}}}, \bibinfo{author}{\bibfnamefont{N.~P.} \bibnamefont{{Topchiev}}}, \bibinfo{author}{\bibfnamefont{O.}~\bibnamefont{{Adriani}}}, \bibinfo{author}{\bibfnamefont{R.~L.} \bibnamefont{{Aptekar}}}, \bibinfo{author}{\bibfnamefont{I.~V.} \bibnamefont{{Arkhangelskaja}}}, \bibinfo{author}{\bibfnamefont{A.~I.} \bibnamefont{{Arkhangelskiy}}}, \bibinfo{author}{\bibfnamefont{L.}~\bibnamefont{{Bergstrom}}}, \bibinfo{author}{\bibfnamefont{E.}~\bibnamefont{{Berti}}}, \bibinfo{author}{\bibfnamefont{G.}~\bibnamefont{{Bigongiari}}}, \bibnamefont{et~al.}, \bibinfo{journal}{arXiv e-prints} \bibinfo{eid}{arXiv:1412.4239} (\bibinfo{year}{2014}), \eprint{1412.4239}.

\bibitem[{\citenamefont{{Boggs}}(2006)}]{Boggs2006}
\bibinfo{author}{\bibfnamefont{S.~E.} \bibnamefont{{Boggs}}}, \bibinfo{journal}{\nar} \textbf{\bibinfo{volume}{50}}, \bibinfo{pages}{604} (\bibinfo{year}{2006}), \eprint{astro-ph/0608532}.

\bibitem[{\citenamefont{{Hunter} et~al.}(2014)\citenamefont{{Hunter}, {Bloser}, {Depaola}, {Dion}, {DeNolfo}, {Hanu}, {Iparraguirre}, {Legere}, {Longo}, {McConnell} et~al.}}]{Hunter2014}
\bibinfo{author}{\bibfnamefont{S.~D.} \bibnamefont{{Hunter}}}, \bibinfo{author}{\bibfnamefont{P.~F.} \bibnamefont{{Bloser}}}, \bibinfo{author}{\bibfnamefont{G.~O.} \bibnamefont{{Depaola}}}, \bibinfo{author}{\bibfnamefont{M.~P.} \bibnamefont{{Dion}}}, \bibinfo{author}{\bibfnamefont{G.~A.} \bibnamefont{{DeNolfo}}}, \bibinfo{author}{\bibfnamefont{A.}~\bibnamefont{{Hanu}}}, \bibinfo{author}{\bibfnamefont{M.}~\bibnamefont{{Iparraguirre}}}, \bibinfo{author}{\bibfnamefont{J.}~\bibnamefont{{Legere}}}, \bibinfo{author}{\bibfnamefont{F.}~\bibnamefont{{Longo}}}, \bibinfo{author}{\bibfnamefont{M.~L.} \bibnamefont{{McConnell}}}, \bibnamefont{et~al.}, \bibinfo{journal}{Astroparticle Physics} \textbf{\bibinfo{volume}{59}}, \bibinfo{pages}{18} (\bibinfo{year}{2014}), \eprint{1311.2059}.

\bibitem[{\citenamefont{{Wu} et~al.}(2014)\citenamefont{{Wu}, {Su}, {Bravar}, {Chang}, {Fan}, {Pohl}, and {Walter}}}]{Wu2014}
\bibinfo{author}{\bibfnamefont{X.}~\bibnamefont{{Wu}}}, \bibinfo{author}{\bibfnamefont{M.}~\bibnamefont{{Su}}}, \bibinfo{author}{\bibfnamefont{A.}~\bibnamefont{{Bravar}}}, \bibinfo{author}{\bibfnamefont{J.}~\bibnamefont{{Chang}}}, \bibinfo{author}{\bibfnamefont{Y.}~\bibnamefont{{Fan}}}, \bibinfo{author}{\bibfnamefont{M.}~\bibnamefont{{Pohl}}}, \bibnamefont{and} \bibinfo{author}{\bibfnamefont{R.}~\bibnamefont{{Walter}}}, in \emph{\bibinfo{booktitle}{Space Telescopes and Instrumentation 2014: Ultraviolet to Gamma Ray}}, edited by \bibinfo{editor}{\bibfnamefont{T.}~\bibnamefont{{Takahashi}}}, \bibinfo{editor}{\bibfnamefont{J.-W.~A.} \bibnamefont{{den Herder}}}, \bibnamefont{and} \bibinfo{editor}{\bibfnamefont{M.}~\bibnamefont{{Bautz}}} (\bibinfo{year}{2014}), vol. \bibinfo{volume}{9144} of \emph{\bibinfo{series}{Society of Photo-Optical Instrumentation Engineers (SPIE) Conference Series}}, p. \bibinfo{pages}{91440F}, \eprint{1407.0710}.

\bibitem[{\citenamefont{{Aramaki} et~al.}(2020)\citenamefont{{Aramaki}, {Adrian}, {Karagiorgi}, and {Odaka}}}]{Grams}
\bibinfo{author}{\bibfnamefont{T.}~\bibnamefont{{Aramaki}}}, \bibinfo{author}{\bibfnamefont{P.~O.~H.} \bibnamefont{{Adrian}}}, \bibinfo{author}{\bibfnamefont{G.}~\bibnamefont{{Karagiorgi}}}, \bibnamefont{and} \bibinfo{author}{\bibfnamefont{H.}~\bibnamefont{{Odaka}}}, \bibinfo{journal}{Astroparticle Physics} \textbf{\bibinfo{volume}{114}}, \bibinfo{pages}{107} (\bibinfo{year}{2020}), \eprint{1901.03430}.

\bibitem[{\citenamefont{{Dzhatdoev} and {Podlesnyi}}(2019)}]{MAST}
\bibinfo{author}{\bibfnamefont{T.}~\bibnamefont{{Dzhatdoev}}} \bibnamefont{and} \bibinfo{author}{\bibfnamefont{E.}~\bibnamefont{{Podlesnyi}}}, \bibinfo{journal}{Astroparticle Physics} \textbf{\bibinfo{volume}{112}}, \bibinfo{pages}{1} (\bibinfo{year}{2019}), \eprint{1902.01491}.

\end{thebibliography}

\end{document}